\documentclass[12pt]{article}

\usepackage{arxiv}
\usepackage{authblk}
\usepackage[utf8]{inputenc} 
\usepackage[T1]{fontenc}    
\usepackage{hyperref}       
\usepackage{url}            
\usepackage{booktabs}       
\usepackage{amsfonts, amsmath}       
\usepackage{nicefrac}       
\usepackage{microtype}      
\usepackage{lipsum}		
\usepackage{graphicx}
\usepackage[super,sort&compress]{natbib}
\setcitestyle{comma,numbers,super,open={},close={}} 
\usepackage{doi}
\usepackage{tikz}
\usetikzlibrary{positioning}
\usepackage[nomarkers,tablesonly]{endfloat}
\usepackage[flushleft]{threeparttable}
\usepackage{multirow}
\usepackage{makecell}
\usepackage{pdflscape}
\usepackage{afterpage}
\usepackage{capt-of}

\usepackage{epstopdf}

\usepackage{placeins}

\usepackage{xcolor, soul} 

\title{A Flexible Multi-Metric Bayesian Framework for Decision-Making in Phase II Multi-Arm Multi-Stage Studies}

\author[1,2]{Suzanne M. Dufault}
\author[3]{Angela M. Crook}
\author[4]{Katie Rolfe}
\author[5]{Patrick P. J. Phillips}
\affil[1]{Division of Biostatistics\\
    Department of Epidemiology and Biostatistics\\
    University of California, San Francisco\\
	San Francisco, CA 94158}
 \affil[2]{UCSF Center for Tuberculosis\\
    University of California, San Francisco\\
	San Francisco, CA 94110}
 \affil[3]{MRC Clinical Trials Unit at UCL\\
        Institute of Clinical Trials \& Methodology\\
        90 High Holborn, 2nd Floor\\
        London WC1V 6LJ}
\affil[4]{GSK\\
        Stevenage, United Kingdom}
\affil[2,5]{Division of Pulmonary and Critical Care Medicine\\
	University of California, San Francisco\\
	San Francisco, CA 94110}

\renewcommand{\headeright}{Manuscript Draft}
\renewcommand{\shorttitle}{Bayesian Decision-Making in Phase II Trials}

\hypersetup{
pdftitle={A flexible multi-metric Bayesian framework for decision-making in Phase II multi-arm studies},
pdfsubject={q-bio.NC, q-bio.QM},
pdfauthor={Suzanne M.~Dufault, Angela~Crook, Katie~Rolfe,Patrick P.J.~Phillips},
pdfkeywords={bayesian, phase II, decision-making, tuberculosis, time to positivity, interim analysis, multi-arm multi-stage},
}

\begin{document}
\maketitle

\begin{abstract}
	We propose a multi-metric flexible Bayesian framework to support efficient interim decision-making in multi-arm multi-stage phase II clinical trials. Multi-arm multi-stage phase II studies increase the efficiency of drug development, but early decisions regarding the futility or desirability of a given arm carry considerable risk since sample sizes are often low and follow-up periods may be short. Further, since intermediate outcomes based on biomarkers of treatment response are rarely perfect surrogates for the primary outcome and different trial stakeholders may have different levels of risk tolerance, a single hypothesis test is insufficient for comprehensively summarizing the state of the collected evidence. We present a Bayesian framework comprised of multiple metrics based on point estimates, uncertainty, and evidence towards desired thresholds (a Target Product Profile) for 1) ranking of arms and 2) comparison of each arm against an internal control. Using a large public-private partnership targeting novel TB arms as a motivating example, we find via simulation study that our multi-metric framework provides sufficient confidence for decision-making with sample sizes as low as 30 patients per arm, even when intermediate outcomes have only moderate correlation with the primary outcome. Our reframing of trial design and the decision-making procedure has been well-received by research partners and is a practical approach to more efficient assessment of novel therapeutics. 
\end{abstract}


\section{Introduction}

Decision-making in phase II clinical trials carries risk and is far from straightforward. While only 18\% of phase II studies establish sufficient evidence to advance a drug into phase III, it seems the wrong drug is often advanced resulting in a failure rate of 50\% of phase III studies\cite{jaki2015multi}. Current approaches are inefficient at differentiating good from poor regimens under phase II settings. Sample sizes tend to be considerably smaller in phase II trials than in phase III. Further, adaptive phase II trials tend to rely on intermediate outcomes for decision-making at interim analyses. While in some disease areas, phase II outcomes are the same as those in phase III\cite{ballantyne2013dolutegravir}, it is common that alternative endpoints are used which may not have perfect correspondence with the primary outcome of interest. In addition to the complications of phase II designs, the typical estimands for decision-making are often suboptimal. Standard approaches in multiarm studies include selecting the $k$ best performing arm(s) or more broadly advancing any arms ``close'' to the best performing arm\cite{jaki2015multi}. A recent extension of Network Meta-Analysis highlighted the pitfalls of basing selection on ranking alone and authors provided recommendations for best practices that ``[consider] not only the magnitude of relative effects but also their uncertainty and overlap of their confidence/credible intervals ''\cite{chaimani2021markov}.  An additional factor for regimen selection in phase II studies is ensuring sufficient evidence has been collected to have confidence that the regimen credibly meets a target product profile with respect to safety, efficacy, and general desirability. Frequentist approaches, such as significance testing and group sequential methods, can advance regimens where there is little to no potential to meet the target product profile\cite{saville2014utility, fisch2015bayesian, pulkstenis2017bayesian}. Bayesian frameworks, using a single or a multi-level framework\cite{fisch2015bayesian, pulkstenis2017bayesian}, have recently been proposed to more directly address the critical question: ``How likely is it that the [target product profile] is [fulfilled] based on my observed data?''\cite{pulkstenis2017bayesian}   

The aim of this paper is to present a Bayesian-supported decision framework which we have developed in the context of a phase II trial with an intermediate endpoint that is not a perfect surrogate and with limited outcome data. We propose a multi-metric approach for 1) ranking of arms and 2) comparison of each arm against a control, using a two-level target product profile. We demonstrate via simulations the potential for de-risking decision-making at interim analyses under a flexible decision framework comprised of metrics incorporating point estimates, estimate variability, and evidence towards desired performance thresholds (i.e., a target product profile).

\section{Proposed Framework}
\subsection{Motivating example\label{subsec:example}}

This decision-making framework is motivated by UNITE4TB, a global public-private partnership with the objective of identifying, in phase IIb trials, new combinations of novel and existing compounds that perform better than the six-month standard of care, HRZE, for the treatment of tuberculosis (TB) when given for four months, thereby supporting evaluation of even shorter durations in a phase IIc trial\cite{boeree2021unite4tb}. The primary clinical outcome in UNITE4TB's PARADIGM4TB-01 trial is the number of unfavorable outcomes (treatment failure, relapse, or re-treatment) occurring within 52 weeks of follow-up. In addition, weekly sputum samples will be collected for twelve weeks post-randomization to monitor the change in time-to-positivity (TTP), defined as ``the time [from inoculation in culture media] it takes for a given sputum sample to yield a positive mycobacteria growth indicator tube culture''\cite{burger2018ttp}. This biomarker, while by no means a validated surrogate endpoint, is available much sooner than the primary endpoint, reflects the potency of the regimen in killing off drug-susceptible TB bacterium\cite{gewitz2021longitudinal}, and is associated with the primary clinical endpoint such that a more potent regimen (one with a steeper change in TTP) is expected to have a lower rate of unfavorable outcomes than a less potent regimen\cite{burger2018ttp,phillips2016limited}. We do not assume either formal individual-level or trial-level surrogacy but instead rely on TTP as a biomarker reflective of trial-level bactericidal behavior.

\subsection{Framework components}
The proposed framework is designed for decision-making during the interim analysis of a multi-stage multi-arm trial and evaluates clinical trial arms comprised of therapeutic regimens based on three critical components described in more detail below: 1) an arm-wise lack of benefit assessment based on the early accumulation of occurrences of the primary endpoint (number of unfavourable outcomes), 2) arm-wise performance based on the recommended decision from the application of the Bayesian multi-level target product profile framework proposed by Pulkstenis, Patra, and Zhang\cite{pulkstenis2017bayesian} as applied to the readily available intermediate endpoint (change in TTP), and 3) the arm-wise relative ranking based on estimation from an appropriate Bayesian model of the intermediate endpoint. Table \ref{tab:performance-measures} displays the specific proposed decision objectives, their triggers, and statistical estimands. 

\begin{table}[!h]
    \centering
    \begin{threeparttable}
        \caption{Proposed quantities for the multi-metric decision-making framework.} \label{tab:performance-measures}
        \begin{tabular}{@{}llp{0.3\textwidth}@{}}
        \toprule
         \textbf{Objective} & \textbf{Trigger}  & \textbf{Statistical Estimand} \\
         \midrule
         1. Arm-wise lack of benefit & High number of observed unfavorable outcomes & No. of unfavorable outcomes $\geq M$  \\ \midrule
         \multirow{3}{*}{2. Arm-wise performance} & NO-GO: Low probability that target value is met & $\textnormal{Pr}_{\theta}(\theta_k \geq \theta_{TV}|X) \leq \tau_{TV}$ \\ \cline{2-3} 
         & Continue: Neither 'NO-GO' nor 'GO' conditions met & \makecell[l]{$\textnormal{Pr}_{\theta}(\theta_k \geq \theta_{TV} |X) > \tau_{TV}$ and \\ $\textnormal{Pr}_{\theta}(\theta_k > \theta_{MAV}|X) \leq 1-\tau_{MAV}$}\\ \cline{2-3} 
         & \makecell[l]{GO: High probability that minimum acceptable value is exceeded \\and at least modest probability that target value might be exceeded} & \makecell[l]{$\textnormal{Pr}_{\theta}(\theta_k \geq \theta_{TV}|X) > \tau_{TV}$ and \\ $\textnormal{Pr}_{\theta}(\theta_k > \theta_{MAV}|X) > 1-\tau_{MAV}$}\\ \midrule
         \multirow{4}{*}{3. Arm-wise relative ranking} & Confidence arm slope is steeper than control & $\textnormal{Pr}_{\theta}(\theta_k > \theta_1 |X)$\\
         & Confidence arm has steepest slope & $\textnormal{Pr}_{\theta}(\theta_k = \theta_{(1)} |X)$\\
         & Confidence arm is in top 2 steepest slopes & $\textnormal{Pr}_{\theta}(\theta_k \in \{\theta_{(1)}, \theta_{(2)}\} |X)$\\
         \bottomrule
    \end{tabular}
    \end{threeparttable}
\end{table}

\begin{figure}
\begin{flushright}
\begin{tikzpicture}[
dashednode/.style={rectangle, draw=black!60, very thick, dashed, minimum size=7mm},
squarednode/.style={rectangle, draw=black!60, very thick, minimum size=5mm},
]
\node[squarednode, align=center]      (q1)               {Has the number of observed\\unfavorable outcomes exceeded\\the pre-determined threshold?};
\node                                 (q0)       [above=0.5cm of q1] {};
\node[squarednode, align=center]      (q2)       [below=of q1] {Is there an acceptable\\probability that it matches\\the target product profile?};
\node[dashednode]                     (q3)       [below=of q2] {Does it rank well?};
\node[black!60!green]                 (go)       [below=of q3] {GO};
\node                                 (temp)     [right= 2cm of q2] {};
\node[black!30!red]                   (stop)     [above= 0.5cm of temp]{STOP};
\node[black!20!yellow]                (continue) [right= 2cm of q3]{Continue};

\draw[->] (q0) -- (q1);
\draw[->] (q1) -- node [right,scale=0.9] {yes} (q2);
\draw[->] (q1) -- node [above,scale=0.9] {no} (stop);
\draw[->] (q2) -- node [right,scale=0.9] {yes} (q3);
\draw[->] (q2) -- node [above,scale=0.9] {no} (stop);
\draw[->] (q2) -- node [right,scale=0.9] {inadequate evidence} (continue);
\draw[->] (q3) -- node [right,scale=0.9] {yes} (go);
\draw[->] (q3) -- node [above,scale=0.9] {no} (continue);
\draw[->] (continue.east) to [out=30,in=45,looseness=2] (q1);
\end{tikzpicture}
\end{flushright}
    \caption{Example flowchart of the decision-making framework applied in a sequential manner. The third component (\textit{Does it rank well?}) is in a dashed-line box as it is only relevant when more than one arm has successfully advanced through the first two decision-making steps.}
    \label{fig:decision-flowchart}
\end{figure}
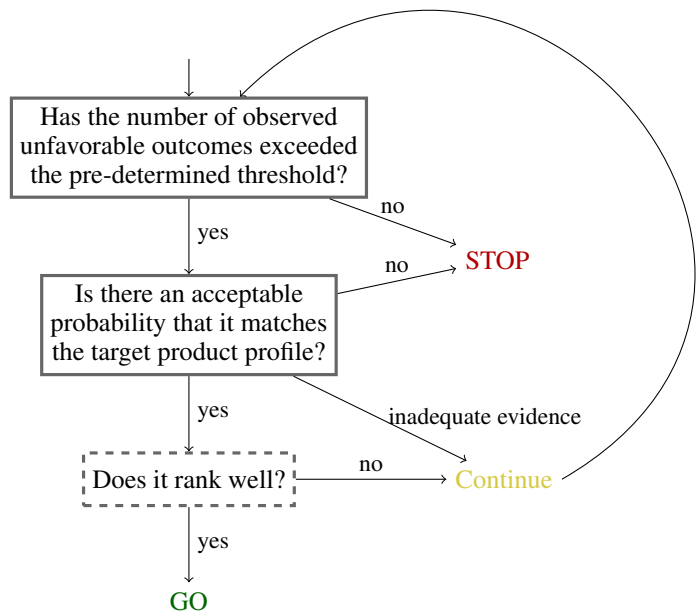

We propose a sequential application of the framework, as it is intuitive and better reflects natural decision-making in terms of predetermined hierarchies of risk tolerance. Figure \ref{fig:decision-flowchart} demonstrates such a stepwise decision-making framework.

\subsubsection{Arm-wise lack of benefit}
Our first objective is to identify and deprioritize sub-optimal arms early. Arms will be flagged for lack of benefit based on whether the number of observed unfavorable outcomes exceeds a set threshold, $M$. While unfavorable outcomes are a definitive clinical outcome and typically used as the primary outcome in phase III studies, they tend to occur after treatment completion\cite{dorman2021four}. As such, there are likely to be very few observed at the time of the interim analysis, thereby limiting our ability to perform statistical analysis. Instead, this can be thought of as an early screening for removal of arms with larger than acceptable anticipated unfavorable event rates. The remaining metrics rely on the intermediate outcome, TTP, as all patients will have TTP data by the time of the interim analysis.

\subsubsection{Arm-wise performance}
Arms are then assessed according to a pre-specified two-level target product profile based on the change in $\log_{10}$(TTP) slope relative to the control slope ($\theta$, expressed as a percentage change). The quantities that must be pre-specified for the target product profile include the ``target value'' or level of efficacy corresponding to solid competitiveness, $\theta_{TV}$ , the ``minimum acceptable value'' or minimal level of acceptable efficacy, $\theta_{MAV}$ , the maximum allowable risk that an arm is issued a NO-GO decision when it has an unequivocal improvement in efficacy, $\tau_{TV}$ , and the maximum allowable risk that an arm is advanced that does not reach the minimal level of acceptable efficacy, $\tau_{MAV}$. 

For each arm $k$, we can then issue a GO, NO-GO, or CONTINUE decision based on the posterior distribution of $\theta_k$. We issue a NO-GO decision if the probability that $\theta_k$ exceeds the target value is sufficiently low ($\Pr_{\theta}(\theta_k\geq\theta_{TV})\leq\tau_{TV}$). We issue a GO decision if the probability that $\theta_k$ exceeds our minimum acceptable value is sufficiently high ($\Pr_{\theta}(\theta_k\geq\theta_{MAV})\geq 1-\tau_{MAV}$) and the probability that $\theta_k$ exceeds our target value is not too low ($\Pr_{\theta}(\theta_k\geq\theta_{TV})> \tau_{TV}$). If neither of these conditions is met, a CONTINUE decision will be issued. Pulkstenis, Patra, and Zhang\cite{pulkstenis2017bayesian} point out that ``there is no universal way to characterize the desired efficacy in the [target product profile]'', and so we refrain from offering general guidance as to specifying the risk and desired parameter values here and instead encourage stakeholders to identify what is appropriate based on historical data and knowledge of the disease area. 

\subsubsection{Arm-wise relative ranking}
Finally, arms are ranked based on a suite of posterior probability estimands targeting their relative ranking and comparison with the control. We also report a credible estimate (median of the Bayesian posterior distribution) for the relative percent-change in $\log_{10}$(TTP) slope as compared to the control, along with a credible interval (confidence level: $1-\alpha$).

\section{Simulation Study Methods}

We describe our simulation study using the the Aims, Data Generation, Estimand, Methods, and Performance Measures (ADEMP) framework outlined by Morris et al\cite{morris2019using}.

\subsection{Aims}

Our overall aim is to evaluate how well the proposed framework can de-risk decision-making around arm selection for multi-arm phase II trials.

\subsection{Data-generating mechanism \label{sec:dgm}}

\textbf{TTP.} The weekly individual-level TTP data is simulated from a parametric linear mixed effects model using the approach described by Arnold et al\cite{arnold2011simulation}. Analysis of longitudinal TTP data from the REMoxTB phase III trial\cite{gillespie2014four} motivated our choice. For individual $i$ and visit $j$, let $T_{ij}$ denote the weeks since randomization at visit $j$. Let $X_i$ denote the assigned treatment arm for individual $i$, $X_i = 1,\ldots,K$ where $X=1$ denotes the control arm. Equation \ref{eq:sim-model} allows for flexibility in individual-level intercepts and slopes.

\begin{align} \label{eq:sim-model}
    \log_{10}(\textnormal{TTP}_{ij}) &= \beta_{0i} + \beta_{1i} T_{ij} +  \sum\limits_{k=2}^K\beta_k\mathbb{I}\{X_i = k\} T_{ij} + e_{ij} 
\end{align}

We pre-specify the random intercept $\beta_{0i} \sim N(\beta_0,\sigma^2_{g_1})$, the random slope $\beta_{1i} \sim N(\beta_1,\sigma^2_{g_2})$, the correlation between the random effects $\rho = \textnormal{Cor}(\beta_{0i}, \beta_{1i})$ and the residual error  $e_{ij} \sim N(0,\sigma^2_e)$. $\mathbb{I}\{\}$ is an indicator function, returning 1 when the condition is true and 0 otherwise. The parameter values used for data-generation are defined in Section A.1 of the Supplemental Material.

\textbf{Unfavorable outcomes.} Individual-level time to unfavorable outcomes, $t_i$, measured from end of treatment, is simulated using a two parameter Weibull proportional hazards model (Equation \ref{eq:weibull}). All individuals are assumed to complete treatment. Assuming there is no loss to follow-up, event times are censored at the end of 52 weeks of post-randomization follow-up if an unfavorable outcome does not occur before. We assume that an individual's hazard of unfavorable outcome depends only on their intervention assignment, not on their individual-level TTP trajectory; correlation between intermediate and final outcomes is therefore induced only at the level of allocated treatment arm. 

\begin{equation}\label{eq:weibull}
\ln h(t_i) = \ln(pt^{p-1}) + \gamma_0 + \sum\limits_{k=1}^K\gamma_k\mathbb{I}\{X_i = k\} + \epsilon_i    
\end{equation}

The Weibull parameters are tuned such that approximately 75\% of unfavorable outcomes occur within the first 13 weeks of post-intervention follow-up\cite{nunn2010timing} (setting scale parameter $p = 0.425$) and such that unfavorable outcomes by the end of follow-up occur according to pre-specified rates. 

\textbf{Interim.} Enrolment dates are randomly assigned such that a rate of ten patients are enrolled per week and randomized to one of five different arms. The first interim analysis occurs one week after complete TTP results are available for the sample size of interest and uses the full TTP data as well as any unfavorable outcome data accumulated up to that point in time. This simulation study only considers the operating characteristics at the first interim analysis.  

\begin{table}[!h]
    \centering
    \begin{threeparttable}
    \caption{Simulation settings for relative percent change in $\log_{10}$(TTP) slope and unfavorable outcome rate. Note, $k = 1$ is the control arm and is used as the comparator.} \label{tab:sim-settings}
    \begin{tabular}{lll}
        \toprule
        Endpoint & Setting  & Conditions (Arm $k$ = 2,3,4,5)\\ 
        \midrule
        \multirow{3}{*}{\makecell{Relative \% TTP Slope (Control: $\theta_1 = 0\%$)\\$\theta_2,\theta_3,\theta_4,\theta_5$}} & 
        No Winners (null) & 0\%, 0\%, 0\%, 0\% \\
         & One Winner & 10\%, 20\%, 30\%, 40\% \\
         & Two Winners & -10\%, 10\%, 35\%, 40\% \\
         & Four Winners & 35\%, 37\%, 39\%, 41\% \\
        \midrule
        \multirow{3}{*}{Unfavorable outcome rates (Control: 5\%)} & All Minimal (null) & 5\%, 5\%, 5\%, 5\% \\
        & All Desirable & 2.5\%, 2.5\%, 2.5\%, 2.5\%\\
        & All Suboptimal & 10\%, 10\%, 10\%, 10\% \\
        & Mixed & 10\%, 5\%, 5\%, 2.5\% \\
        \bottomrule
    \end{tabular}
    \end{threeparttable}
\end{table}

All simulated datasets consist of one control and four novel arms. TTP and unfavorable outcomes were simulated according to the parameterizations in Table \ref{tab:sim-settings}. Contrary to PARADIGM4TB’s 12-week TTP collection plan, the simulated datasets only consist of TTP for 8 weeks post-randomization. This difference is due to changes made to the PARADIGM4TB study design after initial simulations were performed. We consider four settings for TTP slopes representing a null setting (`No Winners') where all TTP slopes are equivalent, evenly spaced slopes with a clear winner (`One Winner'), a mixture of steep and shallow slopes (`Two Winners'), and a setting were all four arms have similarly steep slopes (`Four Winners'). We also consider four settings for unfavorable outcome rates whereby 2.5\% unfavorable outcome is considered desirable, 5\% is considered minimal and 10\% is considered sub-optimal for treatment shortening in the context of a 4-month regimen. All possible combinations of TTP and unfavorable outcome were simulated for each possible sample size in 1,000 simulated datasets. This results in settings where the intermediate and final outcomes were well correlated (steep slopes and low unfavorable outcome rates correspond) and where they were poorly correlated (shallow slopes and low unfavorable outcome rates correspond, and vice versa). Results for any combinations not described here are available in the Supplemental Material and GitHub repository (\url{https://github.com/sdufault15/tb-seamless-design}).

\subsection{Targets of analysis}

The targets of analysis are the arm decision objectives as supported by the framework metrics (Table \ref{tab:performance-measures}). Specifically, we aim to determine whether the framework, when used with standard phase II sample sizes, is sufficient to determine the appropriate arm(s) to flag for lack of benefit or progress, with an acceptable level of risk.

\subsection{Analysis methods}

The weekly $\log_{10}$(TTP) data are analyzed using a Bayesian linear mixed effects model with random intercept and random slope specified at the level of the individual and weakly informative priors. The model formula is reported in the Appendix (Eq. A.1), but echoes that used for data generation (Eq. \ref{eq:sim-model}). Bayesian methods were chosen since they lend themselves to direct probability statements addressing the likelihood of arm success that better facilitate complex decision-making involving non-statisticians
\cite{saville2014utility, spiegelhalter1994bayesian, ashby2006bayesian}.
 Additionally, in this setting, Bayesian methods are desirable because of their ability to handle limit-censoring of the outcome variable\cite{brmsRpackage}. The maximum recommended MGIT incubation time for a sputum sample is 42 days, resulting in a maximum observable TTP value of 42 days and right censoring of TTP values above this limit\cite{burger2018ttp}. While alternative approaches exist to handling right censored outcome variables, likelihood-based approaches have been integrated into standard Bayesian statistical software and are readily available in the setting of non-linear mixed effects models.
 
Unfavorable outcomes are counted at the arm level and compared against count-based thresholds as described in Table \ref{tab:performance-measures}.

Simulations and analyses are performed using R version 4.1.2 (2021-11-01) ``Bird Hippie''\cite{rstudio}. All code necessary to simulate the data, perform the analyses, and recreate the figures presented in this manuscript is available in a GitHub repository maintained by the first author (\url{https://github.com/sdufault15/tb-seamless-design}). Bayesian estimation was performed with the \texttt{brms} package\cite{brmsRpackage,brmsMultilevelModeling}.

\subsection{Performance measures}

To assess the performance of the proposed multi-metric framework (Table \ref{tab:performance-measures}), we examine how each of these metrics can contribute to decision-making when used simultaneously. While a common-sense approach should be taken to guide decision-making, cosidering all available data including safety data, these results are generated under a series of hypothetical, rigid decision-criteria in order to gain intuition into the operating characteristics of the framework. Because the relationship between TTP and unfavorable outcomes is not well understood, we additionally assess the performance of the framework as the correspondence between TTP slope and unfavorable outcomes becomes less well correlated.

We then investigate the performance of each framework component individually. For arm-wise lack of benefit, we examine the rates of deprioritization for desirable, minimal, and sub-optimal arms when the unfavorable outcome threshold is set at fewer than one, two or three unfavorable outcomes by the time of the first interim analysis. Arm-wise performance is evaluated by the proportion of simulations returning ``GO'', ``NO-GO'', and ``Continue'' decisions for an array of the $\log_{10}$(TTP) slopes and sample sizes. For arm-wise ranking, we focus on the proportion of simulations returning posterior probability estimates that favor the arm with the true steepest slope ($\theta_{(1)}$) over the arm with the true second steepest slope ($\theta_{(2)}$), $\Pr_{\theta}(\hat{\theta}_{(1)} = \theta_{(1)}|X) - \Pr_{\theta}(\hat{\theta}_{(2)} = \theta_{(1)} |X)$, in order to identify our ability to differentiate between top performers as the gap in their performance decreases from 10\% to 2\%.

\section{Results}
We first present results from a single simulated trial to illustrate how the framework is implemented (Section \ref{sec:results_single-trial}). Next, we report the performance results from the simulation study for the overall decision-making framework (Section \ref{sec:results_overall-performance}). Finally, we report the performance of each framework component separately (Section \ref{sec:results_separate-framework-components}).

\subsection{Demonstration in a single simulated trial}\label{sec:results_single-trial}

To demonstrate what this framework will look like in practice, we provide the estimated framework metrics for a single simulated trial (Table \ref{tab:results-single-trial}) where TTP has been simulated according to the `1 Winner' setting and unfavorable outcomes were simulated according to the `Mixed' setting (Table \ref{tab:sim-settings}). As such, for Arm 5 the true relative improvement in TTP slope compared to Arm 1 is 40\% and the true unfavorable outcome rate is a desirable 2.5\%. As expected, in the simulated data very few unfavorable outcomes have accrued, suggesting there is not sufficient evidence of lack of benefit to stop any of the arms at this point. Evidence is accumulating that Arm 4 and Arm 5 will meet the target product profile set for TTP; at a sample size of 30 per arm, there is sufficient evidence that these arms are at least as good as the control in terms of their TTP slopes ($\Pr_{\theta}(\theta_{k} > \theta_{MAV}|X) > 99.1\%$) and a strong posterior probability that the slopes are at least 20\% better than the control ($\Pr_{\theta}(\theta_k>\theta_{TV}|X)>79.0\%$). Arm 2 and Arm 3 do not yet have the same strength of evidence suggesting their (in)ability to meet the target product profile. While it is likely that these arms have steeper slopes than control ($\psi_1$), they rank consistently lower than Arm 4 and Arm 5 ($\psi_2,\psi_3$). If resources are available, it would be prudent to continue enrolling participants on these arms and collect more evidence. If resources are not available, the metrics provided by this framework will contribute to the thoughtful evaluation of which arms to continue and which to stop.

\setlength{\tabcolsep}{4pt}

\begin{table}[!h]
    \centering
    \begin{threeparttable}
    \caption{Interim results from a single simulated phase IIb study with thirty patients per arm. The target product profile assumes $\theta_{MAV} = 0\%, \theta_{TV} = 20\%, \tau_{MAV} = \tau_{TV} = 0.025$.}\label{tab:results-single-trial}
    \begin{tabular}{ccccccccccc}
    \toprule
         & & \multicolumn{9}{c}{Interim 1} \\ \cline{3-11}
        arm & Duration & \makecell{No.\\patients} & \makecell{No.\\unfavorable\\outcomes} & \makecell{$\hat{\theta}_{0.5}$ \\ (95\% CI)} & $\Pr_{\theta}(\theta_k>\theta_{MAV}|X)$& $\Pr_{\theta}(\theta_k>\theta_{TV}|X)$ &\makecell{TPP\\decision} & $\hat{\psi}_1$ & $\hat{\psi}_2$ & $\hat{\psi}_3$ \\
        \midrule
        1 & 26 & 30 & 0 & -- & -- & -- & -- & -- & 0.00 & 0.02  \\
        2 & 16 & 30 & 1 &\makecell{11.1\%\\(-12.9\%, 42.2\%)}& 0.81 & 0.26 & Continue & 0.81 & 0.00 & 0.03 \\
        3 & 16 & 30 & 0 &\makecell{23.3\%\\(-1.2\%, 56.9\%)}& 0.97 & 0.59 & Continue & 0.97 & 0.00 & 0.24 \\
        4 & 16 & 30 & 0 &\makecell{31.9\%\\(0.54\%, 68.8\%)}& 0.99 & 0.79 & Go & 0.99 & 0.05 & 0.71\\
        5 & 16 & 30 & 0 &\makecell{55.6\%\\(25.7\%, 95.0\%)}& 1.00 & 0.99 & Go & 1.00& 0.95 & 1.00 \\
    \bottomrule
    \end{tabular}
    \begin{tablenotes}
      \small
      \item Note: $\Pr_{\theta}(\theta_k>\theta_{MAV}|X) = \hat{\psi}_1$ in this setting because $\theta_k > (\theta_{MAV} = 0\%)$ is equivalent to $\theta_k > \theta_1$. When $\theta_{MAV} \neq 0\%$, these metrics will no longer be equivalent.
      \item $\hat{\theta}_{0.5}$ : median estimate of the posterior distribution on the relative \% change in $\log_{10}$(TTP) slope
      \item $\hat{\psi}_1 = \Pr_{\theta}(\theta_k > \theta_1 |X)$, i.e., probability TTP slope for arm $k$ is steeper than control 
      \item $\hat{\psi}_2 = \Pr_{\theta}(\theta_k = \theta_{(1)} |X)$, i.e., probability TTP slope for arm $k$ is the steepest slope
      \item $\hat{\psi}_3 = \Pr_{\theta}(\theta_k \in \{\theta_{(1)}, \theta_{(2)}\} |X)$, i.e., probability TTP slope for arm $k$ is one of the two steepest slopes
    \end{tablenotes}
    \end{threeparttable}
\end{table}


\subsection{Evaluation of the proposed metrics as an overall package}\label{sec:results_overall-performance}

\begin{figure}
    \centering
    \includegraphics[width=\textwidth]{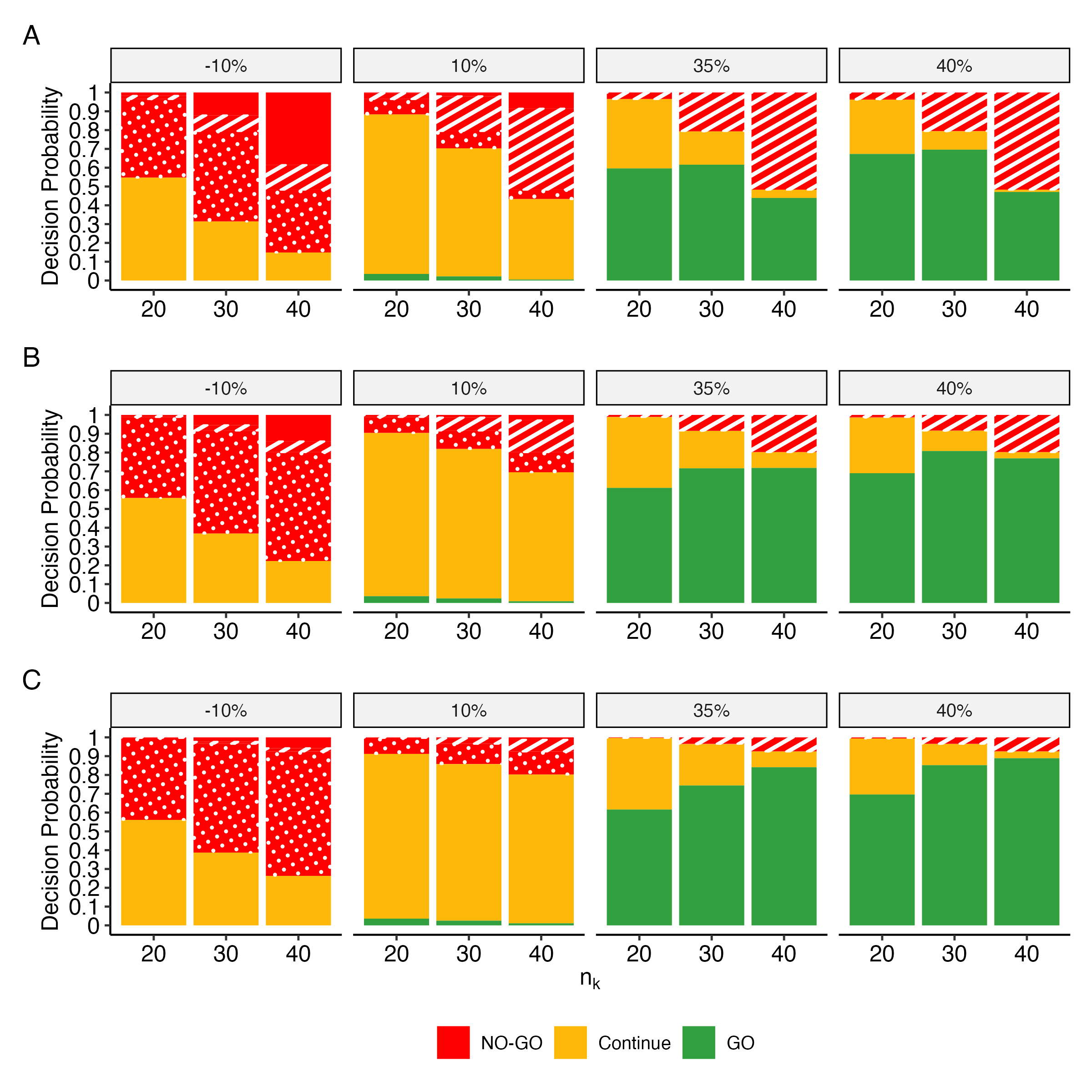}
    \caption{Interim analysis decision probabilities for simulated arms with TTP slope relative to the control as specified in the panels (all corresponding to the `2 Winners' simulation setting) and unfavorable outcome rates of A) 10\% (unfavorable), B) 5\% (minimal), and C) 2.5\% (desirable). Arms are flagged for a GO decision if they meet all the following conditions: fewer than 2 unfavorable outcomes, evidence of meeting the target product profile, and a posterior probability greater than 50\% of ranking in the top two arms. A pattern is applied to differentiate the criterion responsible for NO-GO decisions. Arms are flagged for a NO-GO decision if they experience any of the following conditions: 2 or more unfavorable outcomes (stripes), do not have evidence of meeting the target product profile (dots), or meet both conditions (plain). All arms that do not meet the criteria for a GO or NO-GO decision, receive a ``Continue'' designation. Results are based on 1,000 simulated datasets per setting.}
    \label{fig:full-results}
\end{figure}

Figure \ref{fig:full-results} demonstrates the strength of using each component of the framework in concert to inform decision-making at the time of the interim analysis. This figure only reflects the `2 Winners' TTP simulation setting. The other settings can be observed in the Supplemental Material (Figures A.3--A.5).

First, we examine the operating characteristics for the simulated data where the TTP slope and the unfavorable outcome rate are in correspondence with each other. When the arms have sub-optimal (10\%) unfavorable outcome rates and poor TTP slopes (Fig. \ref{fig:full-results}A, Panel: -10\%), there is a 45.3\% probability of correctly flagging the arm for a ``NO-GO'' decision ($n_k$=20); the probability of correctly flagging these sub-optimal arms increases as sample size increases. Further, none of the suboptimal arms received an erroneous ``GO'' decision, resulting in a false-go rate of 0\% in this setting. When the arms have desirable (2.5\%) unfavorable outcome rates and true TTP slopes above the target product profile target value ($\theta_{TV}$=20\%) (Fig. \ref{fig:full-results}C, Panels: 35\%, 40\%), the probability of a ``NO-GO'' decision is relatively rare. For example, when the TTP slope is 40\% better than control (Fig. \ref{fig:full-results}C, Panel: 40\%), the probability of an erroneous ``NO-GO'' decision is less than 7.5\% ($n_k$ = 40) while the probability of correctly flagging an arm for a ``GO'' decision is at least 88.9\% ($n_k$=40). In other words, if the TTP slope observed at the interim analysis corresponds well with the arm-level unfavorable outcome rates, the framework correctly classifies arm decisions with a level of efficiency not accessible through traditional single-metric means while simultaneously maintaining low error rates in terms of ``false-go'' and ``false-no-go'' rates.

When the correspondence between TTP slope and the primary endpoint diminishes, the effectiveness of the framework diminishes, but does not completely disappear. For example, if TTP slope is a relatively uninformative proxy for the primary endpoint, the framework will reliably flag arms with a ``Continue'' decision (Fig. \ref{fig:full-results}, Panels: 10\%). This allows decision-makers to continue enrolling participants, hopefully to the point where the primary endpoint will have sufficient power for decision-making. If TTP slope has a negative correspondence with the primary endpoint, such that an arm with a sub-optimal unfavorable outcome rate (10\%) returns a relative TTP slope of 40\% at the interim analysis, the risk of making a false-go decision increases to 69.6\% (Fig. \ref{fig:full-results}A, Panel: 40\%, $n_k$=30). Alternatively, if the arm has a desirable unfavorable outcome rate (2.5\%) and a poor relative TTP slope (-10\%), the false-no-go rate increases to the similarly high level of 73.7\% (Fig. \ref{fig:full-results}C, Panel: -10\%, $n_k$=40). However, such extremes are exceptionally unlikely.  

\subsection{Examining each framework component}\label{sec:results_separate-framework-components}

\begin{figure}[!h]
    \centering
    \includegraphics[width = 0.5\textwidth]{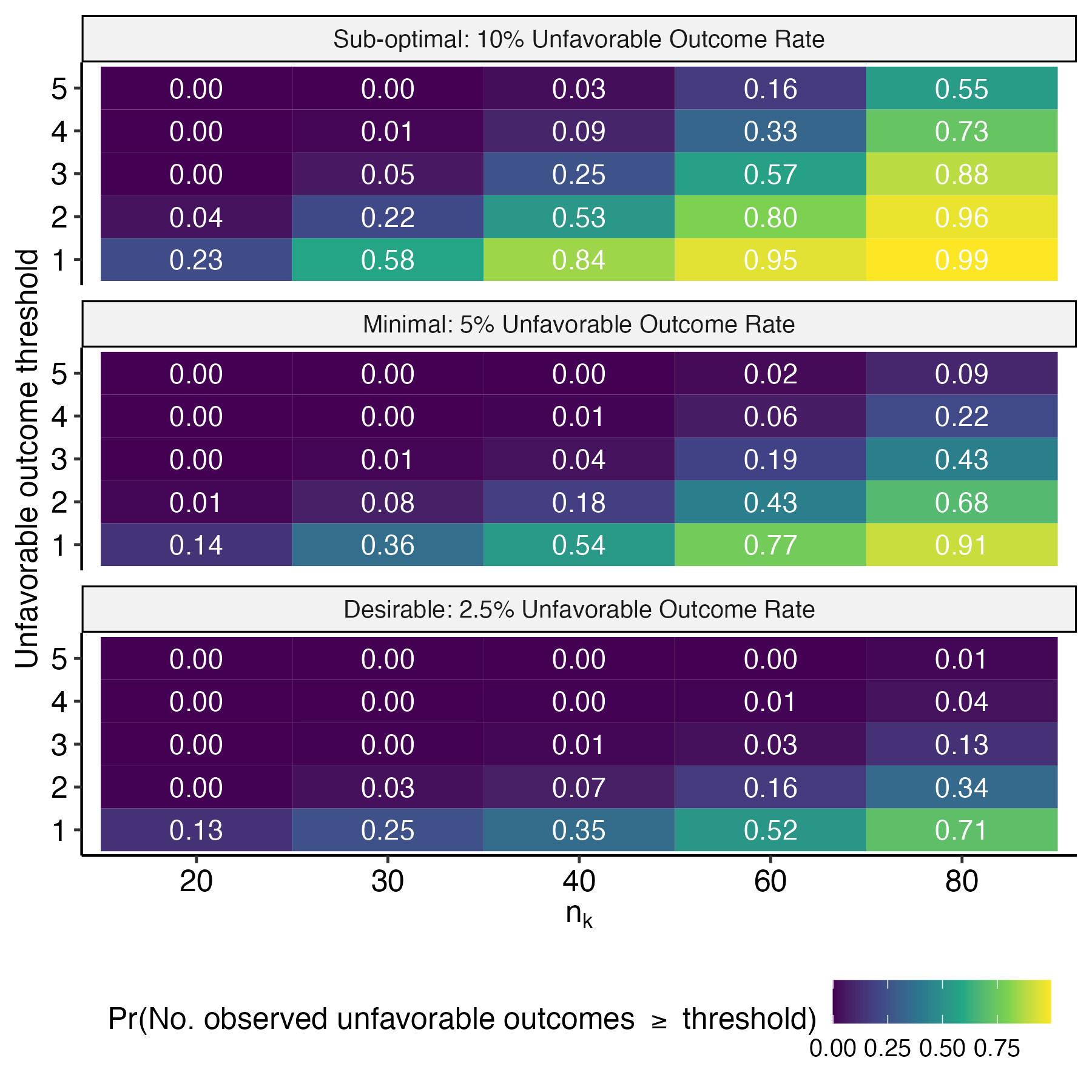}
    \caption{The proportion of simulations where an arm with a given unfavorable outcome rate (panels) would be flagged for deprioritization on the basis of collected unfavorable outcome counts at the first interim analysis given varying sample sizes per arm ($n_k$) and pre-specified unfavorable outcome thresholds. The first interim analysis is triggered by the complete collection of 8 weeks of post-randomization $\log_{10}$(TTP) data on $n_k$ patients per arm. Results are based on the evaluation of 1,000 simulated datasets.}
    \label{fig:unfavorable outcome-alone}
\end{figure}

\textbf{Arm-wise lack of benefit.} Figure \ref{fig:unfavorable outcome-alone} shows the impact of various count-based thresholds for flagging of lack of benefit during the interim analysis. A good decision threshold should result in a high probability for flagging sub-optimal arms and a low probability for desirable arms. It is immediately apparent that decision-makers must be sensitive to the sample sizes considered when pre-specifying the threshold they will use. At a sample size of $n_k = 30$ per arm, an unfavorable outcome threshold of 2 is associated with a 22\% probability of correctly flagging a sub-optimal arm while maintaining a low risk (3\%) of incorrectly flagging a desirable arm for lack of benefit. If the sample size per arm can be increased to $n_k = 40$, the efficiency in flagging sub-optimal arms based solely on early observation of unfavorable outcomes more than doubles (53\%) while maintaining a relatively low risk of flagging a desirable arm (7\%) given the same threshold. 

\begin{figure}[!h]
    \centering
    \includegraphics[width = 0.9\textwidth]{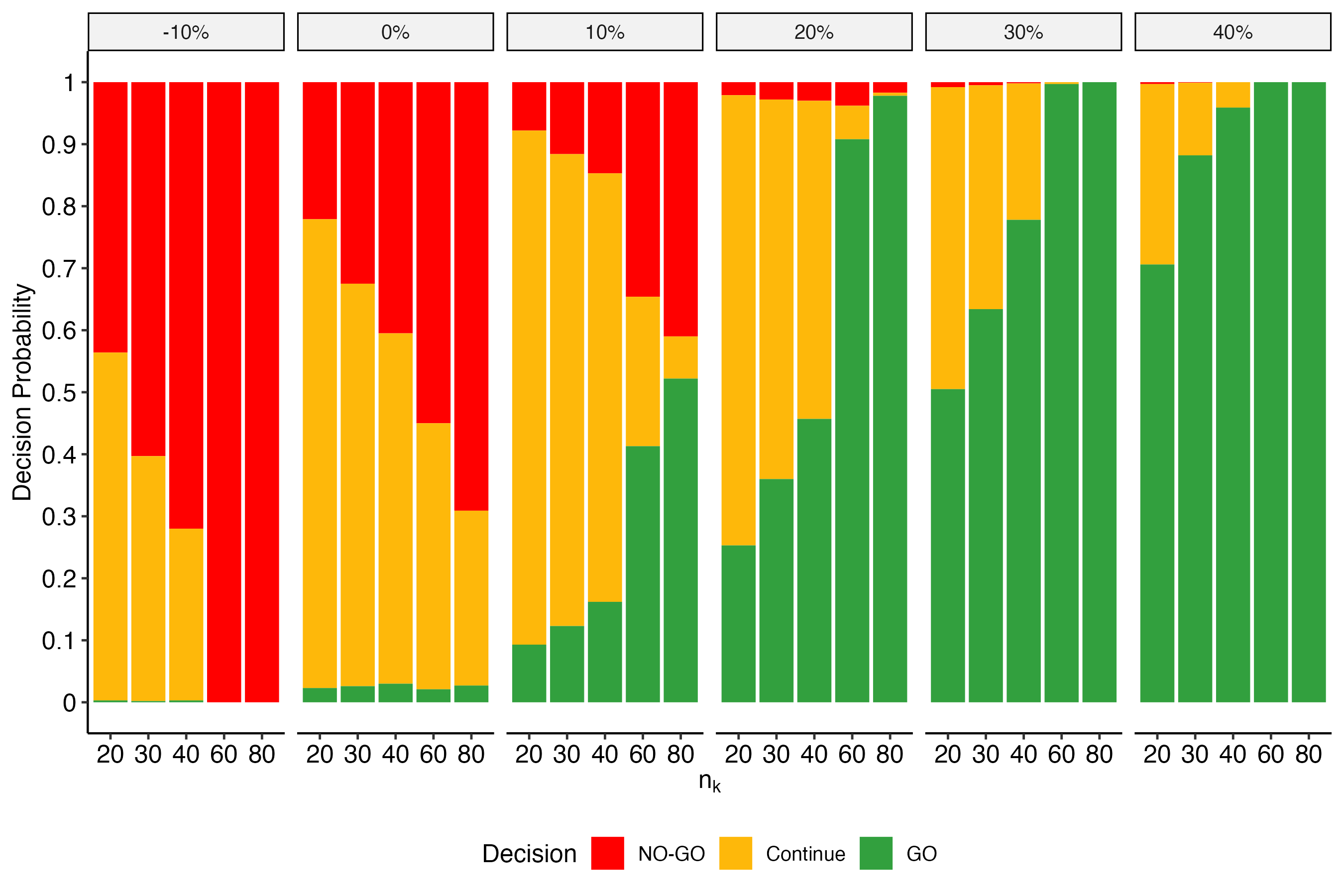}
    \caption{The proportion of trials where an arm with a given percent change in $\log_{10}$(TTP) slope relative to the control (panels) would be assigned a particular decision at the first interim analysis given varying sample sizes per arm ($n_k$). Results are based on the evaluation of 1,000 simulated datasets and assume $\theta_{MAV} = 0\%, \theta_{TV} = 20\%, \tau_{MAV} = \tau_{TV} = 0.025$.}
    \label{fig:tpp}
\end{figure}

\textbf{Arm-wise performance.} Our second step in arm assessment is based on characterizing a two-level target product profile on the $\log_{10}$(TTP) slope. Figure \ref{fig:tpp} displays the impact of assessing arm performance on the basis of the $\log_{10}$(TTP) slope against a multi-level target product profile with the specifed values of $\theta_{MAV} = 0\%, \theta_{TV} = 20\%, \tau_{MAV} = \tau_{TV} = 0.025$. In this setting, an arm with a 10\% poorer slope than the control would be flagged for deprioritization (NO-GO) at least 44\% of the time, even when the sample size is as low as 20 per arm. The probability of advancing (GO) promising arms, those with a $\log_{10}$(TTP) slope 20\% greater than the control, is at least 25\% with a sample size of 20 per arm and increases with increasing sample size. Notably, at a sample size of 40 patients per arm, a promising arm with a $\log_{10}$(TTP) slope 20\% greater than the control is rarely stopped (by design, this proportion hovers around $\tau_{TV}$) and is recommended for early advancement in nearly 50\% of simulations. 

\begin{figure}[!h]
    \centering
    \includegraphics[width = 0.9\textwidth]{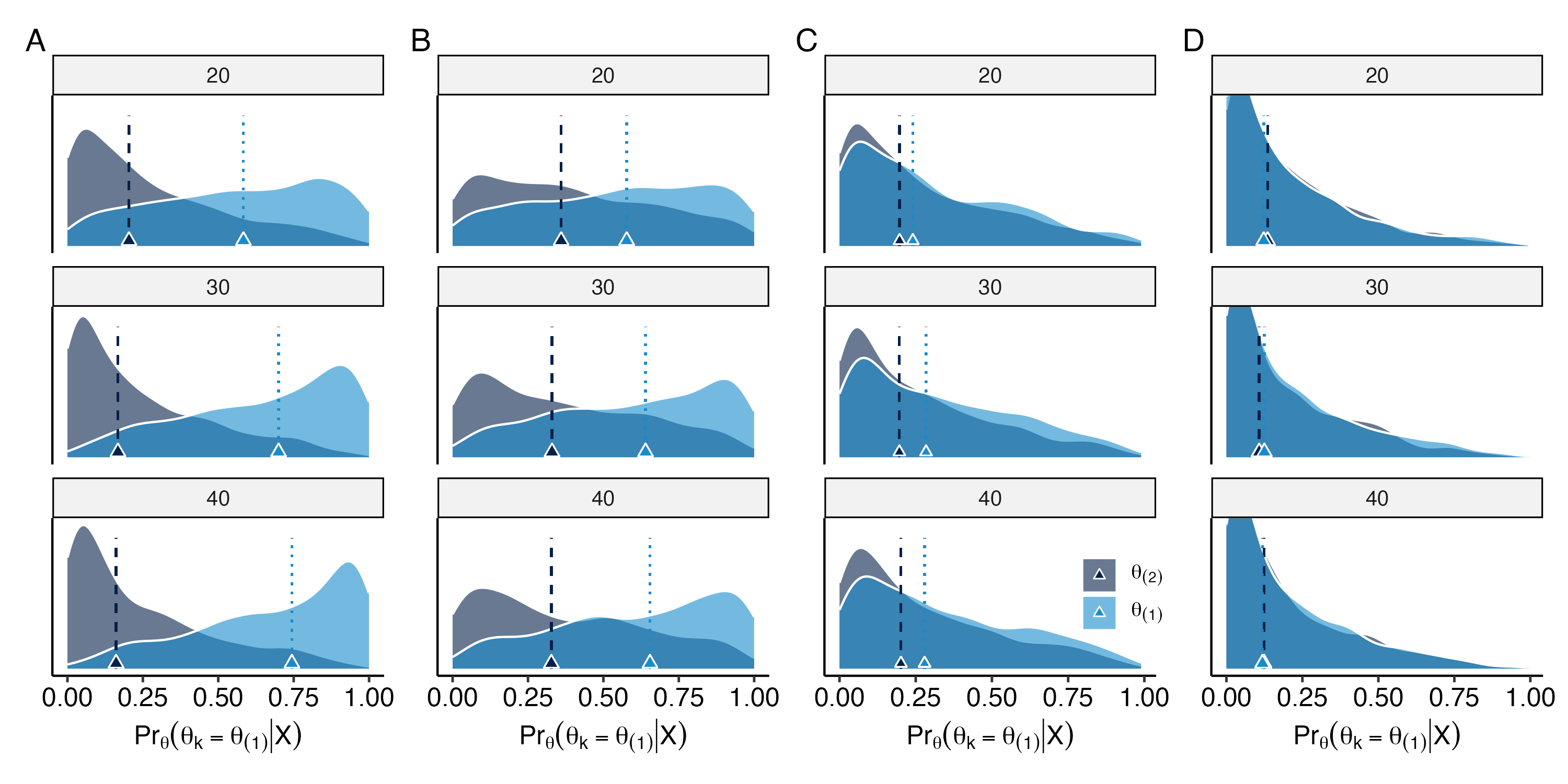}
\caption{Comparison of distributions of posterior probability estimates of whether a given arm has the steepest $\log_{10}$(TTP) slope, $\Pr_{\theta}(\theta_k = \theta_{(1)})$ for the arms with the true steepest $\theta_{(1)}$ and second steepest $\theta_{(2)}$ slopes. Results are shown for differences A) 10\% (`1 Winner'), B) 5\% (`2 Winners'), C) 2\% (`4 Winners'), and D) 0\%, or no difference between arms, (`0 Winners`). Results are based on 1,000 simulated datasets for each sample size (row-wise panels, $n_k$) and TTP condition (column-wise panels). Vertical lines mark the median of the corresponding distributions of posterior probability estimates.}
    \label{fig:posterior-prob-comparison}
\end{figure}

\textbf{Arm-wise ranking.} Figure \ref{fig:posterior-prob-comparison} demonstrates that the ability to properly rank the arm with the true steepest slope depends on sample size and competitiveness of the other arms. For clarity, we have restricted these figures to compare the arms with the true steepest and second steepest slopes in $\log_{10}$(TTP). Each density curve corresponds to the distribution of posterior probability estimates that a given arm is the steepest; ideally, the arm with the true steepest slope ($\theta_{(1)}$, blue curve) would have a posterior probability estimate of 1 in all simulations and the other arms would have posterior probability estimates of 0. Despite uncertainty in estimation in small sample sizes, the posterior probability estimates are often sufficiently higher for the arm with the true steepest slope than for its competitors (median, vertical lines), resulting in a sufficient metric for decision-making. For example, when $\theta_{(1)} - \theta_{(2)} \geq 10\%$ (`1 Winner', Fig. \ref{fig:posterior-prob-comparison}A), a sample size of 30 per arm is sufficient to separate the posterior probability distributions in most simulated datasets. The posterior probabilities associated with ranking are also reliably responsive to the setting of `0 Winners' (Fig. \ref{fig:posterior-prob-comparison}D), suggesting this metric will not return deceptive ranking results when arms are truly similar in terms of TTP slope.  

\section{Discussion}

Decision-making at any point along the clinical trial pathway is an inherent challenge. We have proposed a flexible, multi-metric framework to de-risk decision-making at interim analyses during phase II trials in TB and, with slight adaptation, other disease settings. Our framework combines innovation in both performance evaluation (multi-level target product profile frameworks)\cite{pulkstenis2017bayesian} and arm ranking, and couches all estimation in a readily interpretable Bayesian estimation framework. Using a simulation study, we have demonstrated our proposed framework's suitability to capture critical elements of arm performance even when sample sizes are low. By examining increasingly discordant behavior between the intermediate endpoint used in decision-making and the primary endpoint, we have demonstrated how valuable a multiple metric framework becomes for informed decision-making.

Middle-development TB clinical trials have relied on a handful of commonly used candidate biomarkers (e.g., 14-day EBA, colony forming unit counts, proportion culture negative at 2 months, time to stable culture conversion) as well as  novel biomarkers (e.g., MBLA, RS Ratio, gene signature, PET-CT, sputum LAM) to assess regimen efficacy. The relative utility of the various endpoints remains a topic of debate\cite{gewitz2021longitudinal, phillips2016limited, mitchison1993assessment, diacon2012time, phillips2013evaluation, wallis2013early, bonnett2017comparing}. Our work is based on TTP as the intermediate endpoint as it is the most commonly and readily available outcome in TB trials and appears somewhat promising in terms of trial-level correlation with the primary endpoint. In this setting, we are not using TTP on an individual level to predict or anticipate a single patient's likelihood of cure. Instead, we are assuming that, at the trial-level, the intermediate TTP slope and final outcomes are correlated and that the differences between arms that is observed on TTP is meaningfully correlated with the differences expected in terms of arm performance for the primary endpoint. In the presence of a positive individual level correlation (which may be a plausible assumption for existing drugs\cite{mccallum2022high} and perhaps also for new drugs), we anticipate the operating characteristics of the framework to be even more favorable. As research progresses on this endpoint, general learnings about the relevance of TTP for regimen development can be used to adjust the target and minimum acceptable values.  Our proposed framework, when applied with an appropriate model for the intermediate endpoint, can be extended or adapted to alternative biomarkers, should another option (or the inclusion of additional biomarkers) be of interest to decision-makers.

Bayesian methods for the evaluation of Phase II studies are growing in acceptability\cite{ashby2006bayesian} and have been approved by regulatory agencies as the primary method of analysis\cite{ema2006guidance, fda2018guidance}. One advantage of Bayesian estimation is the ability to explicitly state and incorporate prior information into the estimation procedure. In the setting of TB studies, there is a wealth of knowledge around the standard of care. Ignoring the decades of evidence that has been accumulated is inefficient and, perhaps, unethical when phase II studies are required to keep sample sizes low for equipoise. Though not explored here, future research and applications of this framework should consider the effect of incorporating prior information for the $\log_{10}$(TTP) slope for the standard of care. Following guidance generated by ongoing efforts to incorporate translational pre-clinical and clinical data to improve regimen evaluation (e.g., ACTG RAD-TB), such data sources could also be used to inform reasonable priors on novel regimens as well. Proper incorporation of informative priors should decrease estimator variability in the $\log_{10}$(TTP) slopes, ultimately 1) strengthening the ability to compare novel regimens against the standard of care, 2) improving confidence in ranking, particularly for novel regimens with small relative differences in slope, and 3) result in fewer ``Continue'' categorizations within the target product profile framework. Each of these changes will improve efficiency in the evaluation of regimen performance. Further, it is straightforward to perform sensitivity checks on the impact of the priors and can be an additional tool in guiding decision-makers\cite{gabry2017visualization}.   

One concern with the use of Bayesian methods for the planning and analysis of clinical trials is its inability to strictly control the type I error rate. This is further complicated by our recommendation that the multi-metric framework be applied holistically, upon the close evaluation of all metrics to comprehensively evaluate a study arm's performance and promise. These concerns are worth investigating and future research will evaluate how more complex decision frameworks, such as the one proposed here, can be properly evaluated to limit this risk. One key advantage of our multi-metric framework includes a direct adaptability to decision-makers' level of risk tolerance. Instead of focusing on a strict frequentist type I error, we have shown that this framework has good operating characteristics for prioritizing arms with desirable performance and de-prioritizing sub-optimal arms which directly addresses the objectives of middle-development clinical trials. Further, strict control of the type I error rate may not be the driving determinant in study design for some trial settings. In UNITE4TB-01, this framework can be used to identify which arms advance from phase IIb to phase IIc, a period of further observation where the duration of the arm is also randomized. Evidence generated in this second phase will help to further elucidate which arms (and durations) should be advanced into large, definitive phase III trials. 

In summary, we propose a Bayesian decision framework, building on the two-level target product profile\cite{pulkstenis2017bayesian}, for the setting of multi-arm middle development clinical trials using intermediate endpoints that are not perfect surrogates. We have shown that our flexible multi-metric framework has good operating characteristics and is a practical solution for de-risking drug development. 

\section*{Disclaimer}
This communication reflects the views of the authors and neither IMI nor the European Union and EFPIA are liable for any use that may be made of the information contained herein.

\section*{Conflict of Interest}
UNITE4TB (academia and industry united innovation and treatment for tuberculosis) is a public-private partnership with representation from academic institutions, small- and medium-sized enterprises (SMEs), public organizations, and pharmaceutical companies. All partners of UNITE4TB were provided the opportunity to review a final version of this manuscript for factual accuracy, but the authors are solely responsible for final content and interpretation. Katie Rolfe is employed by and holds shares in GSK. Angela M. Crook and Katie Rolfe are co-leaders of the 'Clinical Trial Design' Work Package within the UNITE4TB consortium.

\section*{Acknowledgments}
This project has received funding from the Innovative Medicines Initiative 2 Joint Undertaking (JU) under grant agreement No 101007873. The JU receives support from the European Union's Horizon 2020 research and innovation programme and EFPIA, Deutsches Zentrum f{\"u}r Infektionsforschung e. V. (DZIF), and Ludwig-Maximilians-Universit{\"a}t M{\"u}nchen (LMU). EFPIA/AP contribute to 50\% of funding, whereas the contribution of DZIF and the LMU University Hospital Munich has been granted by the German Federal Ministry of Education and Research.

Suzanne M. Dufault has received funding from the UCSF Center for Tuberculosis and TB RAMP scholar program (NIH/NIAID R25AI147375).

\FloatBarrier
\newpage
\bibliographystyle{ama}
\bibliography{references.bib}  

\makeatletter\renewcommand\@biblabel[1]{#1.}\makeatother\begin{thebibliography}{10}

\bibitem{jaki2015multi}
Jaki T. Multi-arm clinical trials with treatment selection: what can be gained
  and at what price?.  {\it Clinical Investigation. } 2015;5(4):393--399.

\bibitem{ballantyne2013dolutegravir}
Ballantyne AD, Perry CM. Dolutegravir: First global approval.  {\it Drugs. }
  2013;73(14):1627--1637.

\bibitem{chaimani2021markov}
Chaimani A, Porcher R, Sbidian {\'E}, Mavridis D. A Markov chain approach for
  ranking treatments in network meta-analysis.  {\it Statistics in Medicine. }
  2021;40(2):451--464.

\bibitem{saville2014utility}
Saville BR, Connor JT, Ayers GD, Alvarez J. The utility of Bayesian predictive
  probabilities for interim monitoring of clinical trials.  {\it Clinical
  Trials. } 2014;11(4):485--493.

\bibitem{fisch2015bayesian}
Fisch R, Jones I, Jones J, Kerman J, Rosenkranz GK, Schmidli H. Bayesian design
  of proof-of-concept trials.  {\it Therapeutic Innovation \& Regulatory
  Science. } 2015;49(1):155--162.

\bibitem{pulkstenis2017bayesian}
Pulkstenis E, Patra K, Zhang J. A Bayesian paradigm for decision-making in
  proof-of-concept trials.  {\it Journal of Biopharmaceutical Statistics. }
  2017;27(3):442--456.

\bibitem{boeree2021unite4tb}
Boeree M, Lange C, Thwaites G, \emph{et~al.} UNITE4TB: a new consortium for
  clinical drug and regimen development for TB.  {\it The International Journal
  of Tuberculosis and Lung Disease. } 2021;25(11):886.

\bibitem{burger2018ttp}
Burger DA, Schall R, Chen DG. Robust Bayesian nonlinear mixed-effects modeling
  of time to positivity in tuberculosis trials.  {\it Pharmaceutical
  Statistics. } 2018;17(5):615--628.

\bibitem{gewitz2021longitudinal}
Gewitz AD, Solans BP, Mac~Kenzie WR, \emph{et~al.} Longitudinal Model-Based
  Biomarker Analysis of Exposure-Response Relationships in Adults with
  Pulmonary Tuberculosis.  {\it Antimicrobial Agents and Chemotherapy. }
  2021;65(10):e01794--20.

\bibitem{phillips2016limited}
Phillips PP, Mendel CM, Burger DA, \emph{et~al.} Limited role of culture
  conversion for decision-making in individual patient care and for advancing
  novel regimens to confirmatory clinical trials.  {\it BMC Medicine. }
  2016;14(1):1--11.

\bibitem{dorman2021four}
Dorman SE, Nahid P, Kurbatova EV, \emph{et~al.} Four-month rifapentine regimens
  with or without moxifloxacin for tuberculosis.  {\it New England Journal of
  Medicine. } 2021;384(18):1705--1718.

\bibitem{morris2019using}
Morris TP, White IR, Crowther MJ. Using simulation studies to evaluate
  statistical methods.  {\it Statistics in Medicine. } 2019;38(11):2074--2102.

\bibitem{arnold2011simulation}
Arnold BF, Hogan DR, Colford JM, Hubbard AE. Simulation methods to estimate
  design power: an overview for applied research.  {\it BMC Medical Research
  Methodology. } 2011;11(1):1--10.

\bibitem{gillespie2014four}
Gillespie SH, Crook AM, McHugh TD, \emph{et~al.} Four-month moxifloxacin-based
  regimens for drug-sensitive tuberculosis.  {\it New England Journal of
  Medicine. } 2014;371(17):1577--1587.

\bibitem{nunn2010timing}
Nunn AJ, Phillips PP, Mitchison DA. Timing of relapse in short-course
  chemotherapy trials for tuberculosis.  {\it The International Journal of
  Tuberculosis and Lung Disease. } 2010;14(2):241--242.

\bibitem{spiegelhalter1994bayesian}
Spiegelhalter DJ, Freedman LS, Parmar MK. Bayesian approaches to randomized
  trials.  {\it Journal of the Royal Statistical Society: Series A (Statistics
  in Society). } 1994;157(3):357--387.

\bibitem{ashby2006bayesian}
Ashby D. Bayesian statistics in medicine: A 25 year review.  {\it Statistics in
  Medicine. } 2006;25(21):3589--3631.

\bibitem{brmsRpackage}
Bürkner PC. {brms}: An {R} package for {Bayesian} multilevel models using
  {Stan}.  {\it Journal of Statistical Software. } 2017;80(1):1--28.

\bibitem{rstudio}
{R Core Team} . {\it R: A Language and Environment for Statistical Computing}.
\newblock R Foundation for Statistical ComputingVienna, Austria 2021.

\bibitem{brmsMultilevelModeling}
Bürkner PC. Advanced {Bayesian} multilevel modeling with the {R} package
  {brms}.  {\it The R Journal. } 2018;10(1):395--411.

\bibitem{mitchison1993assessment}
Mitchison DA. Assessment of new sterilizing drugs for treating pulmonary
  tuberculosis by culture at two months.  {\it American Review of Respiratory
  Disease. } 1993;147(4):1062-3.

\bibitem{diacon2012time}
Diacon AH, Maritz J, Venter A, Van~Helden PD, Dawson R, Donald PR. Time to
  liquid culture positivity can substitute for colony counting on agar plates
  in early bactericidal activity studies of antituberculosis agents.  {\it
  Clinical Microbiology and Infection. } 2012;18(7):711--717.

\bibitem{phillips2013evaluation}
Phillips PP, Fielding K, Nunn AJ. An evaluation of culture results during
  treatment for tuberculosis as surrogate endpoints for treatment failure and
  relapse.  {\it PloS one. } 2013;8(5):e63840.

\bibitem{wallis2013early}
Wallis RS, Nacy C. Early bactericidal activity of new drug regimens for
  tuberculosis.  {\it The Lancet. } 2013;381(9861):111--112.

\bibitem{bonnett2017comparing}
Bonnett LJ, Ken-Dror G, Koh GC, Davies GR. Comparing the efficacy of drug
  regimens for pulmonary tuberculosis: meta-analysis of endpoints in
  early-phase clinical trials.  {\it Clinical Infectious Diseases. }
  2017;65(1):46--54.

\bibitem{mccallum2022high}
McCallum AD, Pertinez HE, Chirambo AP, \emph{et~al.} High intrapulmonary
  rifampicin and isoniazid concentrations are associated with rapid sputum
  bacillary clearance in patients with pulmonary tuberculosis.  {\it Clinical
  Infectious Diseases. } 2022.

\bibitem{ema2006guidance}
{European Medicines Agency} . Guideline on clinical trials in small
  populations.  tech. rep. 2006.
\newblock Report No.: CHMP/EWP/83561/2005.

\bibitem{fda2018guidance}
{Food and Drug Administration} . Adaptive designs for clinical trials of drugs
  and biologics: guidance for industry.  tech. rep. 2019.
\newblock Report No.: FDA-2018-D-3124.

\bibitem{gabry2017visualization}
Gabry J, Simpson D, Vehtari A, Betancourt M, Gelman A. Visualization in
  Bayesian workflow.  {\it Journal of the Royal Statistical Society. } 2019.

\end{thebibliography}

\end{document}